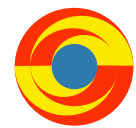

# Heavy Quarkonia and $b\bar{c}$ Mesons in the Cornell Potential with Harmonic Oscillator Potential in the N-dimensional Schrödinger Equation

**M. Abu-Shady**

Department of Mathematics, Faculty of Science, Menoufia University, Shebin El-Kom, Egypt

**Email address:**
abu_shady_1999@yahoo.com

**Abstract:** Heavy quarkonia, $b\bar{c}$ meson, and $c\bar{s}$-meson masses are calculated within the framework of the N-dimensional radial Schrödinger equation. The Cornell potential is extended by including the harmonic oscillator potential. The energy eigenvalues and the corresponding wave functions are calculated in the N-dimensional space using the Nikiforov-Uvarov (NV) method. The energy eigenvalues are obtained in the three-dimensional space. The mass of spectra of charmonium, bottomonium, $b\bar{c}$, and $c\bar{s}$ mesons are calculated. The effect of dimensionality number on the mass of quarkonium is investigated. A comparison with other theoretical approaches is discussed. The obtained results are in good agreement with experimental data. We conclude that the dimensionality number plays an important role in studying the spectra of quarkonium masses. The modified Cornell potential provides a good description of the spectra of quarkonium masses.

## 1. Introduction

The study of heavy quarkonium systems provides good understating for the quantitative test of quantum chromodynamic (QCD) theory and the standard model [1]. There are many techniques that provide us the quantitative and qualitative features about the strong interactions such as the lattice QCD, the QCD sum rules, and other theoretical methods [2]. The Schrödinger equation (SE) plays an important role in the nuclear and subnuclear physics, in particular, the properties of constituents particles and dynamics of their interactions. The investigation of quarkonium systems is widely studied by using the SE. It is known that the exact solutions of the SE are very difficult when the centrifugal potential is included [3]. Thus, there are some methods to find the solution of the SE. In addition, the explicit potential is required to find the explicit solution of the SE.

The methods are widely used for providing the analytic solutions of the SE such as the asymptotic iteration method [4] and the Nikiforov-Uvarov (NU) method [5]. In these methods, the authors use an interaction potential such as the Cornell potential [1, 2, 6] or mixed between the Cornell potential and harmonic oscillator potential [2, 7, 8, 9] or Morse potential [10].

Recently, some authors focus to extend the SE to the higher dimensional which gives more details about the systems under study. Moreover, the energy eigenvalue and wave function are obtained in the lower dimensional space [2].

The aim of this work is to study the N-dimensional of radial SE to obtain the energy eigenvalues and the corresponding wave functions by using the Nikiforov-Uvarov method for the modified Cornell potential, and then apply the present solutions to calculate the heavy quarkonia masses.

The paper is organized as follows: In Sec. 2, the NU method is briefly explained. In Sec. 3, The energy eigenvalues and the corresponding wave functions are calculated in the N-dimensional form. In Sec. 4, the results are discussed. In Sec. 5, the summary and conclusion are presented.

## 2. Theoretical Description of the Nikiforov-Uvarov (NU) Method

In this section, we briefly give the NU method [5] that is used as the technique to solve second-order differential equation which takes the following form

$$\Psi''(s) + \frac{\bar{\tau}(s)}{\sigma(s)}\Psi'(s) + \frac{\tilde{\sigma}(s)}{\sigma^2(s)}\Psi(s) = 0, \quad (1)$$

where $\sigma(s)$ and $\tilde{\sigma}(s)$ are polynomials of maximum second degree and $\bar{\tau}(s)$ is a polynomial of maximum first degree with an appropriate $s = s(r)$ coordinate transformation. To find the particular solution of Eq. (1) by separation of variables, if one deals with the transformation

$$\Psi(s) = \Phi(s)\chi(s), \quad (2)$$

Eq. (1) can be written as in Ref. [1]

$$\sigma(s)\chi''(s) + \tau(s)\chi'(s) + \lambda\chi(s) = 0. \quad (3)$$

where

$$\sigma(s) = \pi(s)\frac{\Phi(s)}{\Phi'(s)}, \quad (4)$$

and

$$\tau(s) = \bar{\tau}(s) + 2\pi(s); \quad \tau'(s) < 0, \quad (5)$$

$$\lambda = \lambda_n = -n\tau'(s) - \frac{n(n-1)}{2}\sigma''(s), n = 0, 1, 2, \ldots \quad (6)$$

$\chi(s) = \chi_n(s)$ is a polynomial of $n$ degree which satisfies the hypergeometric equation, taking the form

$$\chi_n(s) = \frac{B_n}{\rho_n}\frac{d^n}{ds^n}(\sigma^n(s)\rho(s)), \quad (7)$$

where $B_n$ is a normalization constant and $\rho(s)$ is a weight function which satisfies the following equation

$$\frac{d}{ds}\omega(s) = \frac{\tau(s)}{\sigma(s)}\omega(s), \quad \omega(s) = \sigma(s)\rho(s), \quad (8)$$

$$\pi(s) = \frac{\sigma'(s) - \bar{\tau}(s)}{2} \pm \sqrt{(\frac{\sigma'(s) - \bar{\tau}(s)}{2})^2 - \tilde{\sigma}(s) + K\sigma(s)}, \quad (9)$$

and

$$\lambda = K + \pi'(s), \quad (10)$$

the $\pi(s)$ is a polynomial of first degree. The values of $K$ in the square-root of Eq. (9) is possible to calculate if the expressions under the square root are square of expressions. This is possible if its discriminate is zero.

## 3. The Schrödinger Equation and the Cornell Potential with the Harmonic Oscillator Potential

The SE for two particles interacting via symmetric potential in N-dimensional space takes the form as in Ref. [2]

$$[\frac{d^2}{dr^2} + \frac{N-1}{r}\frac{d}{dr} - \frac{L(L+N-2)}{r^2} + 2\mu(E-V(r))]\Psi(r) = 0, \quad (11)$$

where $L, N,$ and $\mu$ are the angular momentum quantum number, the dimensionality number, and the reduced mass for the quarkonium particle, respectively. Setting the wave function $\psi(r) = r^{\frac{1-N}{2}}R(r)$, the following radial SE is obtained

$$[\frac{d^2}{dr^2} + 2\mu(E-V(r)) - \frac{(L+\frac{N-2}{2})^2 - \frac{1}{4}}{2\mu r^2}]R(r) = 0. \quad (12)$$

The Cornell potential plus the harmonic oscillator potential are suggested as in Ref. [2]. Thus $V(r)$ takes the form

$$V(r) = ar - \frac{b}{r} + cr^2 \quad (13)$$

where a, b, and c are arbitrary constants to be determined later. The potential has distinctive features of strong interaction: The confinement and the asymptotic freedom which represent in the first and second terms, respectively. The combined potential of two terms is called Cornell potential. The term $cr^2$ is called the harmonic oscillator potential which the parameter $c$ is proportional to $m$ and $\omega^2$, where $m$ is the mass particle oscillates with frequency $\omega$. This potential plays an important role on the effect of the quarkonium properties as in Refs. [3, 11]. By substituting Eq. (13) into Eq. (12), we obtain

$$[\frac{d^2}{dr^2} + 2\mu(E - ar + \frac{b}{r} - cr^2) - \frac{(L+\frac{N-2}{2})^2 - \frac{1}{4}}{2\mu r^2}]R(r) = 0. \quad (14)$$

Let us assume that $r = \frac{1}{x}$ and $r_0$ is a characteristic radius of the meson. Then the scheme is based on the expansion of $\frac{1}{x}$ in a power series around $r_0$, i.e. around $\delta = \frac{1}{r_0}$ in $x$ space (for details, see Ref. [3]). Eq. (14) takes the following form

$$[\frac{d^2}{dx^2} + \frac{2x}{x^2}\frac{d}{dx} + \frac{2\mu}{x^4}(-A + Bx - C_1 x^2)]R(x) = 0, \quad (15)$$

where, $A = -\mu(E - \frac{3a}{\delta} - \frac{6c}{\delta^2})$, $B = \mu(\frac{3a}{\delta^2} + \frac{8c}{\delta^3} + b)$, and $C_1 = \mu(\frac{a}{\delta^3} + \frac{c}{\delta^4} + \frac{(L+\frac{N-2}{2})^2 - \frac{1}{4}}{2\mu})$. By comparing Eq. (15) and Eq. (1), we find $\bar{\tau}(s) = 2x$, $\sigma(s) = x^2$, and $\tilde{\sigma}(s) = 2\mu(-A + Bx - C_1 x^2)$. Hence, the Eq. (15) satisfies the conditions in Eq. (1). By following the NU method that mentioned in Sec. 2, therefore

$$\pi = \pm\sqrt{(K+2C_1)x^2 - 2Bx + 2A}. \quad (16)$$

The constant $K$ is chosen such as the function under the square root has a double zero, i.e. its discriminant $\Delta = 4B^2 - 8A(K+2C_1) = 0$. Hence,

$$\pi = \pm \frac{1}{\sqrt{2A}}(2A - Bx). \quad (17)$$

Thus,

$$\tau = 2x \pm \frac{2}{\sqrt{2A}}(2A - Bx). \quad (18)$$

For bound state solutions, we choose the positive sign in the above equation so that the derivative is given

$$\tau' = 2 - \frac{2B}{\sqrt{2A}} \quad (19)$$

By using Eq. (10), we obtain

$$\lambda = \frac{B^2}{2A} - 2C_1 - \frac{B}{\sqrt{2A}}, \quad (20)$$

and Eq. (6), we obtain

$$\lambda_n = -n\left(2 - \frac{2B}{2\sqrt{A}}\right) - n(n-1). \quad (21)$$

From Eq. (6); $\lambda = \lambda_n$. The energy eigenvalue of Eq. (14) in the N-dimensional space is given

$$E_{nL}^N = \frac{3a}{\delta} + \frac{6c}{\delta^2} - \frac{2\mu(\frac{3a}{\delta^2} + b + \frac{8c}{\delta^3})^2}{[(2n+1) \pm \sqrt{1 + \frac{8\mu a}{\delta^3} + 4((L+\frac{N-2}{2})^2 - \frac{1}{4}) + \frac{24\mu c}{\delta^4}}]^2} \quad (22)$$

One obtains the energy eigenvalue in Ref. [1] by taking $c = 0$ and $N = 3$. One obtains the energy eigenvalue in Ref. [3] by taking $a = 0$ and $N = 3$. By following, the steps in the section 2. The radial of the wave function of Eq. (14) takes the following form

$$R_{nL}(r) = C_{nL} r^{-\frac{B}{\sqrt{2A}}-1} e^{\sqrt{2A}r} (-r^2 \frac{d}{dr})^n (r^{-2n+\frac{2B}{\sqrt{2A}}} e^{-2\sqrt{2A}r}), \quad (23)$$

For detail, see Ref. [3]. $C_{nL}$ is the normalization constant that is determined by $\int |R_{nL}(r)|^2 d\mathbf{r} = 1$. In Eq. (23), we note that the radial wave function does not explicitly depend on the number of dimensions. Hence, $\int |R_{nL}(r)|^2 dr = 1$ remains unchanged, one obtains the wave function in Ref. [3], by taking $a = 0$ and wave function in Ref. [1], by taking $c = 0$.

## 4. Results and Discussion

In this section, we calculate spectra of the heavy quarkonium system such charmonium and bottomonium mesons that have the quark and antiquark flavor, the mass of quarkonium is calculated in 3-dimensional ($N = 3$). So we apply the following relation as in Refs. [1, 2]

$$M = 2m + E_{nL}^{N=3} \quad (24)$$

where $m$ is bare quark mass for quarkonium. By using Eq. (22), we can write Eq. (24) as follows:

$$M = 2m + \frac{3a}{\delta} + \frac{6c}{\delta^2} - \frac{2\mu(\frac{3a}{\delta^2} + b + \frac{8c}{\delta^3})^2}{[(2n+1) \pm \sqrt{1 + \frac{8\mu a}{\delta^3} + 4L(L+1) + \frac{24\mu c}{\delta^4}}]^2} \quad (25)$$

In Table (1), the charmonium mass is calculated by using Eq. (25) in comparison with other theoretical methods and the experimental data. We select the positive sign in Eq. (25) in the present calculations as in Ref. [3]. In addition, the free parameters of the present calculations $a, b, c$, and $\delta$ are fitted with experimental data and Eq. (25) as in Ref. [3]. In addition, quark masses $c$ and $b$ are obtained from Ref. [1]. In comparison with results in Ref. [2], the authors calculated the charmonium mass using the asymptotic iteration method (AIM) and the same potential as in Eq. (13). We note that the present results are improved in comparison with results in the AIM and are in good agreement with experimental data. Al-Jamel and Widyan [3] calculated charmonium and bottomonium masses, in which they employed the Coulomb potential plus quadratic potential. We have some advantages in comparison with the results in Ref. [3], in which NU method is used.

**Table 1.** *Mass spectra of charmonium (in GeV) ($m_c = 1.209$ GeV [1], $a = 5.9530 \times 10^{-2}$ GeV$^2$, $\delta = 0.27557$ GeV, $b = 0.5$ and $c = 2.4277 \times 10^{-2}$ GeV$^3$).*

| State | Present work | [2] | [3] | [7] | [1] | [12] | N=4 | Exp. [13] |
|---|---|---|---|---|---|---|---|---|
| 1S | 3.096 | 3.078 | 3.096 | 3.096 | 3.096 | 3.078 | 3.360 | 3.096 |
| 1P | 3.259 | 3.415 | 3.433 | 3.433 | 3.255 | 3.415 | 3.673 | - |
| 2S | 3.686 | 4.187 | 3.686 | 3.686 | 3.686 | 3.581 | 3.698 | 3.686 |
| 1D | 3.511 | 3.752 | 3.767 | 3.770 | 3.504 | 3.749 | 3.895 | - |
| 2P | 3.779 | 4.143 | 3.910 | 4.023 | 3.779 | 3.917 | 3.827 | 3.773 |
| 3S | 4.037 | 5.297 | 3.984 | 4.040 | 4.040 | 4.085 | 3.966 | 4.040 |
| 4S | 4.263 | 6.407 | 4.150 | 4.358 | 4.269 | 4.589 | 3.986 | 4.263 |
| 2D | 4.093 | - | - | 3.096 | - | 3.078 | 4.170 | 4.159 |

The first, in Ref. [3], the potential is the particular case of the present potential when to take $a = 0$. So we can obtain energy eigenvalues and the corresponding wave functions of Ref. [3] by taking $a = 0$ and $N = 3$. The second, the present results are improved in comparison with results in Ref. [3] and are in good agreement with experimental data. The third, we find that an increase in the dimensionality number increases the charmonium masses as in Table 1. The dimensionality number is not found in Ref. [3]. In comparison with Ref. [7], the authors used the same potential as in Eq. (13) and used NU method, we have some advantages that the SE is extended to the higher dimensional space. Hence, the energy of quarkonium meson and wave function are extended to the N-dimensional space, leading the expression of the energy and wave function are different. Hence, the present results are improved in comparison with Ref. [7] and are in good agreement with experimental data. Also, we add further investigations by calculating $b\bar{c}$ meson and $c\bar{s}$ meson masses in the present work as in Tables 3 and 4. In comparison with Ref. [1], the author used the Cornell potential only. Therefore, the Cornell potential is the particular case at $c = 0$ in the present potential. The present results are improved in comparison with results in Ref. [1] and are in good agreement with experimental data, in which NU method is used. In comparison with Ref. [12], we note that the present results are improved and are in good agreement with experimental data.

**Table 2.** Mass spectra of bottomonium (in GeV) ($m_b = 4.823$ GeV [1], $a = 0.17653$ GeV$^2$, $\delta = 0.38119$ GeV, $b = 1.569$ and $c = 6.169 \times 10^{-3}$ GeV$^3$).

| State | Present work | [2] | [3] | [8] | [1] | [12] | N=4 | Exp. [13] |
|---|---|---|---|---|---|---|---|---|
| 1S | 9.460 | 9.510 | 9.460 | 9.460 | 9.460 | 9.510 | 9.610 | 9.460 |
| 1P | 9.612 | 9.862 | 9.840 | 9.811 | 9.619 | 9.862 | 10.022 | |
| 2S | 10.023 | 10.627 | 10.023 | 10.023 | 10.023 | 10.038 | 10.072 | 10.023 |
| 1D | 9.849 | 10.214 | 10.140 | 10.161 | 9.864 | 10.214 | 10.205 | |
| 2P | 10.111 | 10.944 | 10.160 | 10.374 | 10.114 | 10.390 | 10.269 | |
| 3S | 10.361 | 11.726 | 10.280 | 10.355 | 10.355 | 10.566 | 10.306 | 10.355 |
| 4S | 10.580 | 12.834 | 10.420 | 10.655 | 10.567 | 11.094 | 10.344 | 10.580 |

**Table 3.** Mass spectra of $b\bar{c}$ meson (in GeV) ($m_b = 4.823$ GeV, $m_c = 1.209$ GeV, $a = 0.2$ GeV$^2$, $\delta = 0.54103$ GeV, $b = 1.2$ and $c = 0.04$ GeV$^3$). (In Eq. (25), $2m = m_b + m_c$).

| State | Present work | [14] | [15] | [16] | N=4 | Exp. [13] |
|---|---|---|---|---|---|---|
| 1S | 6.277 | 6.349 | 6.264 | 6.270 | 6.355 | 6.277 |
| 1P | 6.666 | 6.715 | 6.700 | 6.699 | 6.883 | - |
| 2S | 7.042 | 6.821 | 6.856 | 6.835 | 6.878 | - |
| 2P | 7.207 | 7.102 | 7.108 | 7.091 | 7.161 | - |
| 3S | 7.384 | 7.175 | 7.244 | 7.193 | 8.035 | - |

In Table 2, the bottomonium mass is calculated. We note that all states of bottomonium meson are improved in comparison with other theoretical methods. In addition, the present results are in good agreement with experimental data. We find that an increase in the dimensionality number increases mass spectra of bottomonium. In Table 3, the $b\bar{c}$ meson mass is calculated. We find that the 1S state closes with experimental data. The experimental data of the other states are not available. Hence, the theoretical predictions using the present method and other theoretical are displayed. We note that the present results of the $b\bar{c}$ meson mass are in good agreement in comparison with Refs. [14,15,16].

**Table 4.** Mass spectra of $c\bar{s}$-meson (in GeV) (($m_c = 1.628$, $m_s = 0.419$) GeV [19], $a = 0.01$ GeV$^2$, $\delta = 0.58492$ GeV, $b = 1.9138$ and $c = 6.1704 \times 10^{-2}$ GeV$^3$) (In Eq. (25), $2m = m_c + m_s$).

| | | Potential | | | | |
|---|---|---|---|---|---|---|
| State | Present work | Power | Screened | Phenomenological | N=4 | Exp. |
| 1S | 1.968 | 1.9724 | 1.9685 | 1968 | 2.3001 | 1968.3 [18] |
| 1P | 2.566 | 2.540 | 2.7485 | 2.566 | 2.742 | - |
| 2S | 2.709 | 2.6506 | 2.8385 | 2.815 | 2.797 | 2.709 [19] |
| 3S | 2.932 | 2.9691 | 3.2537 | 3.280 | 2.967 | - |
| 1D | 2.857 | - | - | - | 2.934 | 2.859 [19] |

In Table 4 shows the mass spectra of $D(c\bar{s})$-meson. We note that the state 1S is in good agreement with experimental data. Also, the states of 2S and 1D are in good agreement with experimental data measured by the LHCb as in Ref. [19]. The experimental data of the other states are not available. Hence, the theoretical predictions using the present method and other theoretical are displayed, in which the present results are in good agreement with the power potential, the screened potential, and the phenomenological potential in Ref. [17]. In Table 4 shows that the states of $c\bar{s}$-meson mass increase by increasing dimensionality number at $N = 4$ due to increase in the corresponding binding energies of spectra of $D(c\bar{s})$-meson.

## 5. Summary and Conclusion

In the present work, the energy eigenvalues and the wave functions are obtained in the N-dimensional form by solving the N-radial Schrödinger equation. The NU method is employed as the technique to solve the Schrödinger equation, in which the Cornell potential is extended by including the harmonic oscillator term. We find:

I-The effect of dimensionality number is studied, we find that increasing dimensionality number increases charmonium, bottomonium, $b\bar{c}$ meson masses, and $c\bar{s}$-meson masses as in Tables (1, 2, 3, and 4). This effect is not considered in many works such as in Refs. [1,2,3,8].

II-In the N-dimensional space, the energy eigenvalues and wave functions are obtained and they coincide with the recent works such as in Refs. [1,3] when dimensionality number is reduced to $N = 3$.

III -A comparison is studied with other theoretical methods, in which the advantages of the present potential are displayed. In addition, the obtained results are in good agreement with experimental data.

Therefore, we conclude that the dimensionality number plays an important role in studying the spectra of quarkonium masses. The modified Cornell potential provides a good description of the spectra of quarkonium masses in comparison with other theoretical methods and is in good agreement with experimental data. We hope to extend this work for further investigations of other characteristics of quarkonium.

## Acknowledgement

The author thanks Prof. S. M. Kuchin for useful comments that supported this work.